\documentclass[12pt]{article}
\pdfoutput=1
\usepackage{amsmath,amsfonts,graphicx,color,bbm,tikz,bm,setspace}
\usepackage{comment}
\usepackage[nosort]{cite}
\usepackage{subfigure}
\usepackage{qcircuit}
\usepackage{multirow}
\usetikzlibrary{calc,positioning}
\usetikzlibrary{patterns,arrows,decorations.pathreplacing}
\usepackage{caption}
\usepackage{ulem}
\tikzset{>=stealth}
\usepackage{lipsum}
\usepackage{tabularx}
\usepackage{longtable}
\usepackage{arydshln}
\usepackage{fullpage}
\usepackage{hyperref}
\usepackage{bbold}
\usepackage[english]{babel}

\newtheorem{remark}{Remark}[section]

\usepackage{listings}
\usepackage{tcolorbox}

\newcommand{\mathsym}[1]{{}}
\newcommand{\unicode}[1]{{}}

\makeatletter\@addtoreset{equation}{section}\makeatother

\newcommand{\be}{\begin{equation}}
\newcommand{\ee}{\end{equation}}
\def\beq{\begin{equation}}
\def\eeq{\end{equation}}
\newcommand{\bea}{\begin{eqnarray}}
\newcommand{\eea}{\end{eqnarray}}

\newcommand{\ket}[1]{{\left| {#1} \right>}}

\renewcommand{\title}[1]{\vbox{\center\LARGE{#1}}\vspace{3mm}}
\renewcommand{\author}[1]{\vbox{\center{#1}}\vspace{3mm}}

\newcommand{\email}[1]{\vbox{\center\tt#1}\vspace{3mm}}

\begin{document}

\begin{center}
{\large {\bf Non-hermitian integrable systems from constant non-invertible solutions of the Yang-Baxter equation }}

\author{Somnath Maity,$^a$ Pramod Padmanabhan,$^a$ Vladimir Korepin$^{b}$}

{$^a${\it School of Basic Sciences,\\ Indian Institute of Technology, Bhubaneswar, 752050, India}}
\vskip0.1cm
{$^b${\it C. N. Yang Institute for Theoretical Physics, \\ Stony Brook University, New York 11794, USA}}

\email{somnathmaity126@gmail.com, pramod23phys@gmail.com, vladimir.korepin@stonybrook.edu}

\vskip 0.5cm 

\end{center}


\abstract{
\noindent 
We construct invertible spectral parameter dependent Yang-Baxter solutions ($R$-matrices) by Baxterizing constant non-invertible Yang-Baxter solutions. The solutions are algebraic (representation independent). They are constructed using supersymmetry (SUSY) algebras. The resulting $R$-matrices are regular leading to local non-hermitian Hamiltonians written in terms of the SUSY generators. As particular examples we Baxterize the $4\times 4$ constant non-invertible solutions of Hietarinta leading to nearest-neighbor Hamiltonians. On comparing with the literature we find that two of the models are new. Apart from being non-hermitian, many of them are also non-diagonalizable with interesting spectrums. With appropriate representations of the SUSY generators we obtain spin chains in all local Hilbert space dimensions.

}

\tableofcontents 

\section{Introduction}
\label{sec:Introduction}
A systematic method to construct quantum integrable systems is through solving the Yang-Baxter equation \cite{YangCN1967,BAXTER1972193}. The resulting quantum and statistical physics models \cite{Baxter1982ExactlySM,Korepin1993QuantumIS,jimbo1990yang} are solved with the different variations of the algebraic Bethe ansatz \cite{Takhtadzhan_1979,slavnov2019algebraicbetheansatz,eckle2019models}. These operator equations contain spectral parameters and are highly non-linear and overdetermined. They can get very difficult to solve especially when the size of the matrix operators increase. A promising method to produce spectral parameter dependent solutions is called {\it Baxterization}. It was introduced by V. F. R. Jones \cite{Jones1990} as a procedure to turn braid group representations into spectral parameter dependent Yang-Baxter solutions. This was demonstrated for the famous Jones representation \cite{Jones:1985dw,kauffman-knots} built out of Temperley-Lieb algebras \cite{Temperley:1971iq}. Quotients of the braid group such as Hecke algebras and the Birman-Murakami-Wenzl (BMW) algebras are also Baxterized using similar methods. Several other techniques for Baxterization can be found in \cite{ge-baxterization,li1993yang,ge-Baxterization-91,kulish-baxterization,ZHANG1991625}. A mathematical interpretation of Baxterization is as follows. Consider a representation of the braid group generators. Then we can ask the question if elements of the braid group algebra will continue to satisfy a braid type relation. The answer is that it sometimes can if we also include spectral parameters. As far as we know there is no universal answer to this question. There is also no unique way to Baxterize a given braid group representation. It is also not true that only braid group generators can be Baxterized (See \cite{Crampe_2016} for the Baxterization of other braid-like algebras.). 

More recently there have been works trying to exhaust all regular solutions of the $4\times 4$ type \cite{vieira2018solving,de2019classifying,Corcoran:2023zax,garkun2024newspectralparameterdependentsolutions} without Baxterization. These use differential and analytic properties of the $R$-matrices to find solutions. Such non-algebraic [representation dependent] approaches could possibly exhaust [classify] regular solutions in a particular dimension of the local Hilbert space. However application of these methods in dimension 3 and above is computationally quite expensive.

In the physics literature braid group representations are also known as constant Yang-Baxter solutions. They do not contain spectral parameters. They have been classified in the $4\times 4$ case by Jarmo Hietarinta \cite{HIETARINTA-PLA,hietarinta1993-JMP-Long}. Their algebraic realizations was only recently studied in \cite{MSPK-Hietarinta}. This helps us obtain the analogous higher dimensional representations. It also helps us to understand generalizations of the Jones representation through the construction of braid group generators using partition algebras \cite{martin1994temperley,martin1996structure,halverson2004partitionalgebras}. Given the profound impact of the Jones polynomial on theoretical physics \cite{witten-QFTjones}, it is important to study the consequences of the knot/link polynomials associated to these new braid group representations.

The spectral parameter dependent Yang-Baxter solutions constructed with the above methods come from the Baxterization of invertible braid group generators. It is then natural to ask:
\begin{enumerate}
    \item Can we Baxterize non-invertible constant Yang-Baxter solutions?
    \item Will these lead to integrable models that are local?
\end{enumerate}
In this work we will answer these questions in the affirmative by using algebraic methods to generate regular solutions $\left[R_{ij}(u=0)=P_{ij}\right]$ with additive spectral parameters. They give rise to integrable Hamiltonians with nearest-neighbor interactions. The local terms are naturally non-hermitian, by which we mean that they are non-hermitian without the need for any complex parameters. They can potentially describe open quantum spin chains or out-of-equillibrium spin chains.

The results are organized in a pedagogical manner. The different forms of the Yang-Baxter equation are explained in Sec. \ref{sec:YBE}. This symmetries of these equations and the definition of an equivalence class of solutions are also explained here (Sec. \ref{subsec:symmetries-YBE}). Following this the algebraic derivation of the spectral parameter dependent solutions are detailed in Sec. \ref{sec:algSolutions}.
Here we also show how supersymmetry (SUSY) algebras provide a natural realization of these solutions in Sec. \ref{subsec:SUSYrealizations}. We apply our methods to the 12 classes of $4\times 4$ constant non-invertible solutions of Hietarinta. The non-hermitian Hamiltonians in each class is derived and compared with known results in the literature in Sec. \ref{sec:4x4-nhHamiltonians}. We find 9 inequivalent classes of nearest-neighbor Hamiltonians in the $4\times 4$ case as summed up in \ref{subsec:Summary}. Among these 9 classes, 2 of them are new to the best of our knowledge. In Sec. \ref{sec:features} the scope of our methods is explained. We show how to derive higher dimensional solutions using $\mathbb{Z}_2$-gradings of $\mathbb{C}^n$ in Sec. \ref{subsec:higherdimensions}. This is followed by a brief analysis of the spectrums of the non-hermitian Hamiltonians obtained in the previous sections in Sec. \ref{subsec:spectrumstudy}. In particular we look at when these Hamiltonians can be diagonalized and when they cannot. We study one of them in more detail to see the interesting features of such non-hermitian Hamiltonians. We end with a few future directions in Sec. \ref{sec:conclusion}.

\section{Yang-Baxter equation}
\label{sec:YBE}
We begin with a review of the Yang-Baxter equation in its different forms.
In the process the notation and terminology we will follow in the rest of the paper is also fixed. The Yang-Baxter equation (YBE) comes in two forms: with or without spectral parameters. The latter is also called the constant YBE. It coincides with the braid relation that appears in the study of braid groups and in low dimensional topology. The former is sometimes referred to as the quantum YBE. It is studied in the context of integrable systems in physics.

\subsection{Constant YBE}
\label{subsec:cYBE}
The constant form is given by\footnote{Note the departure in choice of notation for the Yang-Baxter operator. The Physics literature uses $R$ for the same \cite{HIETARINTA-PLA}. We reserve the use of $R$ for the spectral parameter dependent case. This notation is also different from the braid group and knot theory literature where $\sigma$ or $b$ is preferred over $Y$.}
\begin{eqnarray}\label{eq:YBE-constant-braided}
    Y_{12}Y_{23}Y_{12} = Y_{23}Y_{12}Y_{23}.
\end{eqnarray}
This is an operator equation on $V\otimes V\otimes V$ with $V$ a local Hilbert space. The operator $Y_{ij}$ acts non-trivially on $V_i\otimes V_j$ and trivially (as identity) on all the other sites. The solutions presented in this paper are algebraic which means they are independent of the choice of the dimension of $V$. In general they work for $V\simeq\mathbb{C}^N$ when the appropriate representation is chosen for $Y$. However our focus in later sections will be on $N=2$ or the qubit case.

The constant form given in \eqref{eq:YBE-constant-braided} is also called the {\it braided form} as the index structure of the equation resembles that of the braid relations satisfied by the braid group generators $\sigma$, 
$$  \sigma_i\sigma_{i+1}\sigma_i = \sigma_{i+1}\sigma_i\sigma_{i+1}~;~\sigma_i\equiv\sigma_{i,i+1}.$$
Henceforth we will call \eqref{eq:YBE-constant-braided} as the constant braided YBE (cbYBE). In contrast to this we can obtain an equivalent equation, with a different index structure, called the constant non-braided YBE (cnbYBE),
\begin{eqnarray}\label{eq:YBE-constant-nonbraided}
    \tilde{Y}_{12}\tilde{Y}_{13}\tilde{Y}_{23} = \tilde{Y}_{23}\tilde{Y}_{13}\tilde{Y}_{12}.
\end{eqnarray}
This is the {\it non-braided} form of the constant YBE. The operators $Y$ and $\tilde{Y}$ are related by 
\begin{eqnarray}
    \tilde{Y} = P\cdot Y,
\end{eqnarray}
where $P$ is the permutation operator satisfying the relations
\begin{eqnarray}
    P_iP_{i+1}P_i = P_{i+1}P_iP_{i+1}~;~P^2=\mathbb{1}~;~P_iP_j = P_jP_i,~~|i-j|>1,
\end{eqnarray}
of the permutation group. The cnbYBE appears more in the physics literature due to its utility in deriving integrable systems. Thus this is the standard form of the YBE familiar to the integrability community. We will sometimes refer to the non-braided form as the standard form.

\subsection{Spectral parameter dependent YBE}
\label{subsec:spYBE}
Following the notation used in the constant case, the braided and non-braided forms of the spectral parameter dependent YBE (spYBE) are
\begin{eqnarray}
    & R_{12}(u,v)R_{23}(u,w)R_{12}(v,w) = R_{23}(v,w)R_{12}(u,w)R_{23}(u,v),    & \label{eq:YBE-spec-braided} \\
    & \tilde{R}_{12}(u,v)\tilde{R}_{13}(u,w)\tilde{R}_{23}(v,w) = \tilde{R}_{23}(v,w)\tilde{R}_{13}(u,w)\tilde{R}_{12}(u,v), & \label{eq:YBE-spec-nonbraided}
\end{eqnarray}
respectively. The spectral parameters $u$, $v$ and $w$ are in general complex. The operator $\tilde{R}$ is related to $R$ via the permutation operator $P$ as in the constant case. 

We will denote the equations \eqref{eq:YBE-spec-braided} and \eqref{eq:YBE-spec-nonbraided} as spbYBE and spnbYBE respectively. In the form written, the spectral parameters appear in the non-additive form. Some of the standard examples of integrable systems are obtained from the $R$-matrix obeying the spYBE in the additive or difference form. This implies that the $R$-matrix becomes 
$$  R_{ij}(u,v) \equiv R_{ij}(u-v). $$
Such $R$-matrices only depend on the difference of the spectral parameters.

\subsection{Symmetries of the YBE}
\label{subsec:symmetries-YBE}
Certain transformations of a given Yang-Baxter solution resulting in another solution constitute symmetries of the YBE. They come in continuous and discrete forms. The former is generated by an invertible matrix $Q$,
\begin{eqnarray}\label{eq:cYBE-gauge}
    Y_{ij}\rightarrow \kappa~Q_iQ_jY_{ij}Q_i^{-1}Q_j^{-1}, 
\end{eqnarray}
for the constant solutions and $Q(u_i)$, and
\begin{eqnarray}\label{eq:spYBE-gauge}
    R(u_i,u_j)_{ij}\rightarrow \kappa(u_i, u_j)~Q_i(u_i)Q_j(u_j)R_{ij}(u_i,u_j)Q_i(u_i)^{-1}Q_j(u_j)^{-1}
\end{eqnarray}
for the solutions with complex spectral parameters $u$ that depend on the indices. Here $\kappa$ is a complex constant or $\kappa(u,v)$ is a complex scalar function of the spectral parameters. Note that this leaves the non-additive form of the YBE  \eqref{eq:YBE-spec-braided} and \eqref{eq:YBE-spec-nonbraided} invariant.  
As these transformations are local they are also known as {\it gauge transformations} in earlier works.
\begin{remark}
    The $R$-matrices equivalent to a regular $R$-matrix are also equivalent. However, two spectral parameter dependent $R$ matrices in the same equivalence class in general result in two different Hamiltonians. For example the regular solutions $R(u,v)$ and $\check{R}(u,v)$ related by the similarity transform $Q(u)\otimes Q(v)$ 
    $$ \check{R}_{ij}(u,v) = \kappa(u,v)Q_i(u)Q_j(v)R_{ij}(u,v)Q_i^{-1}(u)Q_j^{-1}(v),  $$
    result in the local Hamiltonians
    \begin{eqnarray}
        & \mathfrak{h}_{ij} = R^{-1}_{ij}(u,u) \frac{dR_{ij}(u,v)}{du}\Biggr|_{u=v} = P_{ij} \frac{dR_{ij}(u,v)}{du}\Biggr|_{u=v},   &  \nonumber \\
        & \check{\mathfrak{h}}_{ij} = Q_i(u)Q_j(u)\mathfrak{h}_{ij}Q_i^{-1}(u)Q_j^{-1}(u) + \frac{d\ln{\kappa(u,v)}}{du}\Biggr|_{u=v}         &  \nonumber \\
        &+ \left[\dot{Q}_i(u)Q(u)_i^{-1}-Q(u)_i^{-1}\dot{Q}_i(u) \right],         &
    \end{eqnarray}
    respectively. Here $\dot{Q}(u)=\frac{dQ(u)}{du}$ and we have used the facts that $R_{ij}(u,u)=R_{ij}^{-1}(u,u)=P_{ij}$ and $\check{R}^{-1}_{ij}(u,u) = \frac{1}{\kappa(u,u)}P_{ij}$. The term in $\left[\cdot\right]$ vanishes on a closed chain when $Q^{-1}$ commutes with $\dot{Q}$. The other term depending only on the scalar function $\kappa(u,u)$ is an overall constant. In such cases it is enough to check for a local similarity transform between Hamiltonians to determine if they fall into the same equivalence class. We will use this condition to compare the local Hamiltonians obtained in this work with those in the literature. It should be noted that this definition of equivalence between local Hamiltonians does not necessarily mean that the two Hamiltonians are physically equivalent as they could still represent two different phases of matter. 
\end{remark}

The discrete symmetries of the YBE are global transformations. The number of such discrete symmetries depend on the dimension of the representation in which we are working in. We will write down these symmetries for the case when the local Hilbert space dimension is 2 or the qubit representation. We will point out the changes when the dimension increases.

In the qubit case there are three discrete symmetries:
\begin{eqnarray}
    & Y \rightarrow Y^T\equiv Y^{(I)}~~~\textrm{Discrete - I},    & \label{eq:YBE-discrete-I} \\
    & Y \rightarrow \left[ X\otimes X\right]Y\left[ X\otimes X\right]\equiv Y^{(II)}~~~\textrm{Discrete - II},   & \label{eq:YBE-discrete-II} \\
    & Y \rightarrow PYP\equiv Y^{(III)}~~~\textrm{Discrete - III}.   & \label{eq:YBE-discrete-III}
\end{eqnarray}
Here $T$ denotes the transpose and $X$ denotes the first Pauli matrix. Similar equations also hold for the spectral parameter dependent solutions $R(u,v)$. The first discrete transformation can be generalized to the adjoint as well. The third discrete symmetry rearranges the indices in the tensor product space leaving the YBE invariant. Both the first and third discrete symmetries continue to hold in the same form in higher dimensions as well.

The second discrete symmetry arises from a relabeling of the basis of the local Hilbert space $\mathbb{C}^2$. As the dimension is 2 this symmetry forms a $\mathbb{Z}_2$ group. In higher dimensions there will be more possibilities for relabeling the basis elements. These are generated by the shift operators in dimension $n$, that generates $\mathbb{Z}_n$. They generalize the Pauli $X$ matrices in higher dimensions. Thus for dimension $n$, we have $n-1$ such second discrete symmetries. This makes the total number of discrete symmetries in dimension $n$ as $n+1$.

The different YBE solutions related by these continuous and discrete symmetries are defined to fall into a single equivalence class. This implies that two solutions of the YBE are seen to be distinct if they are not related by any of these symmetries or any combination of them. We will use this definition to identify inequivalent solutions.

\section{Algebraic solutions}
\label{sec:algSolutions}
Constant Yang-Baxter solutions can be turned into spectral parameter dependent solutions by a process termed Baxterization \cite{Jones1990}. This process shows that when $Y$, the generator of the braid group, solves the cbYBE, the spectral parameter dependent form spbYBE can be solved by an element of the braid group algebra. We will now see some simple solutions of both the braided and non-braided spectral parameter dependent YBE's \eqref{eq:YBE-spec-braided} and \eqref{eq:YBE-spec-nonbraided} obtained by Baxterizing the constant solutions $Y$.

Consider the ansatz
\begin{eqnarray}\label{eq:ansatz-braided}
    R_{ij}(u) = \mathbb{1} + \rho(u)~Y_{ij}.
\end{eqnarray}
Here $\rho(u)$ is an unknown function of the complex spectral parameter $u$. This solves the spbYBE when either one of the following conditions hold:
\begin{enumerate}
    \item When $Y^2=0$, then $\rho(u)=cu$, with $c$ a constant. This solves the additive form of the YBE.
    \item When $Y^2=\eta~Y$, then $\rho(u) = \frac{e^{cu}-1}{\eta}=\frac{\sinh{cu}+\cosh{cu}-1}{\eta}$. This also solves the additive form of the YBE.
    \item When $Y^2=\eta~\mathbb{1}$, then again $\rho(u)=cu$ and the ansatz solves the braided form of the YBE.
\end{enumerate}
In this work we will only construct solutions using the first two conditions. It is evident that in both these cases the constant solution $Y$ is non-invertible. Thus we have 
\begin{eqnarray}
    & R_{ij}(u) = \mathbb{1} + cu~Y_{ij}~;~R_{ij}^{-1}(u) = \mathbb{1} - cu~Y_{ij}~~;~~Y^2=0,   & \label{eq:R-braided-1} \\
    & R_{ij}(u) = \mathbb{1} + \frac{e^{cu}-1}{\eta}~Y_{ij}~;~  R_{ij}^{-1}(u) = \mathbb{1} + \frac{e^{-cu}-1}{\eta}~Y_{ij}~~;~~Y^2=\eta~Y. \label{eq:R-braided-2}   &
\end{eqnarray}
The local Hamiltonians are derived from the non-braided forms of these solutions $$ \tilde{R}_{ij}(u) = P_{ij}R_{ij}(u).$$ These are given by
\begin{eqnarray}\label{eq:H-braided-1}
    H = \sum\limits_j \tilde{R}^{-1}_{j,j+1}(u) \tilde{R}'_{j,j+1}(u)\Biggr|_{u=0}= \sum_j R^{-1}_{j,j+1}(u) R'_{j,j+1}(u)\Biggr|_{u=0} = c \sum\limits_j Y_{j,j+1},
\end{eqnarray}
for \eqref{eq:R-braided-1} and
\begin{eqnarray}\label{eq:H-braided-2}
     H = \sum\limits_j R^{-1}_{j,j+1}(u) R'_{j,j+1}(u)\Biggr|_{u=0} = \frac{c}{\eta} \sum\limits_j Y_{j,j+1}
\end{eqnarray}
for \eqref{eq:R-braided-2} respectively. 

Next we consider an ansatz for the non-braided form of the spectral parameter dependent YBE \eqref{eq:YBE-spec-nonbraided},
\begin{eqnarray}\label{eq:R-nonbraided-1}
    \tilde{R}_{ij}(u) = P_{ij} + \rho(u)~\tilde{Y}_{ij}.
\end{eqnarray}
This satisfies spnbYBE when\footnote{There exists another possibility where $P\tilde{Y}P\tilde{Y}=\eta~\mathbb{1}$, with $\eta$ a complex constant. We do not consider this possibility in this work.} 
$$ P\tilde{Y}P\tilde{Y} = 0~\textrm{and}~\rho(u) = cu. $$
Note that this constraint on $Y$ implies that $\tilde{Y}P\tilde{Y}P=0$ as well.
The inverse is given by 
\begin{eqnarray}
    \tilde{R}^{-1}_{ij}(u) = P_{ij}- cu~P_{ij}\tilde{Y}_{ij}P_{ij}.
\end{eqnarray}
The corresponding Hamiltonian is then given by
\begin{equation}\label{eq:H-nonbraided-1}
    H = \sum\limits_j \tilde{R}^{-1}_{j,j+1}(u) \tilde{R}'_{j,j+1}(u)\Biggr|_{u=0} = c \sum\limits_j P_{j,j+1} \tilde{Y}_{j,j+1}.
\end{equation}
The Yang-Baxter solutions and the corresponding local Hamiltonians constructed thus far are completely algebraic, in the sense that they are valid for all choices of the local Hilbert space $\mathbb{C}^n$. The form of the constant solutions $Y$, are also not specified. They only have to satisfy either of the following three Baxterization conditions
\begin{eqnarray}\label{eq:3-conditions}
    & Y^2=0~~~\textrm{Condition I}, & \nonumber \\ 
    & Y^2=\eta~Y~~~\textrm{Condition II}, & \nonumber \\  
    & P\tilde{Y}P\tilde{Y}=0~~~\textrm{Condition III}.
\end{eqnarray}
We now look at some simple constant Yang-Baxter solutions $Y$ fulfilling these conditions before we move on to a more systematic treatment with the help of Hietarinta's $4\times 4$ constant non-invertible solutions. This will provide some algebraic realizations of $Y$ that is independent of the choice of the dimension of the local Hilbert space $V$.

\subsection{Examples}
\label{subsec:examples-algebraic}
Consider the following factorized ansatz for $Y$\footnote{Such solutions are special cases of those found in \cite{padmanabhan2024yangbaxtersolutionscommutingoperators}.}
\begin{eqnarray}\label{eq:Y-AA}
    Y_{ij} = S_iS_j.
\end{eqnarray}
This can satisfy either of the three conditions listed in \eqref{eq:3-conditions}. The first two conditions on $Y$, namely a nilpotent $Y$ or a projector $Y$ is achieved by requiring that $S$ is also nilpotent or squares to itself up to a factor, respectively. All these choices are regardless of the dimension of the representation of $S$, making these solutions algebraic. In both cases the operator $Y$ satisfies both the cbYBE \eqref{eq:YBE-constant-braided} and cnbYBE \eqref{eq:YBE-constant-nonbraided}. These choices gives us local Hamiltonians of the form 
$$ H\sim \sum\limits_j~S_jS_{j+1}. $$
These are seemingly simple Hamiltonians with the local neighboring terms commuting with each other. However it should be noted that these Hamiltonians are non-hermitian. Moreover some of them are also non-diagonalizable as shall we see later with specific examples. 

When $S$ is nilpotent, the third condition on $Y$, namely $P\tilde{Y}P\tilde{Y}=0$, is also satisfied. Thus in this case we have another local Hamiltonian given by
$$ H\sim \sum\limits_j~P_{j,j+1}S_jS_{j+1}.  $$

Another simple solution of the factorized form is given by 
\begin{eqnarray}\label{eq:Y-AB}
    Y_{ij} = S_iT_j.
\end{eqnarray}
There is no apriori relation between $S$ and $T$. This implies that they need not satisfy the constant forms of the YBE as well. However when either $S$ or $T$ is nilpotent they will satisfy the condition that $Y^2=0$. In this case the operator $Y$ satisfies both the constant forms of the YBE. The local Hamiltonian is then 
$$ H\sim \sum\limits_j~S_jT_{j+1}. $$
On the other hand if they are both proportional to projectors then they only satisfy the cnbYBE, giving rise to the same kind of local Hamiltonian. The neighboring terms of this local Hamiltonian in both cases do not commute with each other. This makes this example more non-trivial than the ones obtained from \eqref{eq:Y-AA}.

This solution for $Y$ can also fulfill the $P\tilde{Y}P\tilde{Y}=0$ condition when either one of $ST=0$ or $TS=0$ is satisfied. This gives rise to the local Hamiltonian
$$ H\sim \sum\limits_j~P_{j,j+1}S_jT_{j+1}.  $$
As in the case of \eqref{eq:Y-AA}, all of these solutions are algebraic [representation independent].

\subsection{SUSY realizations } 
\label{subsec:SUSYrealizations}
SUSY algebras provide a natural realization of nilpotent operators and projectors. So we define this next. Consider the following relations generated by the supercharge $q$ and its adjoint $q^\dag$,
\begin{eqnarray}\label{eq:SUSY-relations}
    & q^2=0,~~(q^\dag)^2=0,     & \\
    & \{q, q^\dag\} = qq^\dag + q^\dag q = b + f =h.&
\end{eqnarray}
Here $h$ is the SUSY Hamiltonian. It is a sum of the projectors ($b$, $f$) to the bosonic and fermionic sectors. The projector conditions follow from the fact that
\begin{eqnarray}
    bq=qq^\dag q= q~;~fq^\dag=q^\dag qq^\dag = q^\dag.
\end{eqnarray}
These relations form the $\mathcal{N}=2$ SUSY algebra. It describes SUSY quantum mechanics \cite{Witten1981DynamicalBO,Cooper_1995}. We can also view it as generators of a $\mathbb{Z}_2$-grading of a Hilbert space.

Using these generators it is easy to see that 
$$ S=q~\textrm{or}~S=q^\dag$$
realizes nilpotent $S$'s.
On the other hand when 
$$ S=b~\textrm{or}~S=f,$$
we obtain projector $S$'s. These operators help construct the constant $Y_{ij}$ operator $S_iS_j$. They can satisfy either of the three conditions set in \eqref{eq:3-conditions}. The corresponding Hamiltonians are either given by \eqref{eq:H-braided-1} or \eqref{eq:H-nonbraided-1}.

The operator $Y_{ij}=S_iT_j$ is Baxterized when either one of $ST=0$ or $TS=0$ is satisfied. This is easily realized using the SUSY generators\footnote{For solutions of the generalized Yang-Baxter equation using similar realizations see \cite{padmanabhan2019quantum}.} as 
\begin{eqnarray}
    ST = 0 \begin{cases}
        S=q,~T=b \\
        S=q^\dag,~T=f \\
        S=b,~T=q^\dag \\
        S=f,~T=q.\\
        S=b,~T=f,
    \end{cases}
\end{eqnarray}
The $TS=0$ yields the same choices when $S$ and $T$ are interchanged in the above equation. The Hamiltonians in this case are of the form \eqref{eq:H-nonbraided-1}.

All of the above SUSY realizations are algebraic or independent of the size of the representation of the SUSY algebra. For the $\mathbb{C}^2$ representation the SUSY generators take the form
\begin{eqnarray}\label{eq:SUSY-C2}
    & q=\begin{pmatrix}
        0 & 1 \\ 0 & 0
    \end{pmatrix}~;~q^\dag=\begin{pmatrix}
        0 & 0 \\ 1 & 0
    \end{pmatrix},   &  \nonumber \\
     & b=qq^\dag=\begin{pmatrix}
        1 & 0 \\ 0 & 0
    \end{pmatrix}~;~f=q^\dag q=\begin{pmatrix}
        0 & 0 \\ 0 & 1
    \end{pmatrix}.   &
\end{eqnarray}
It is clear that these matrices perform a $\mathbb{Z}_2$-grading of the Hilbert space $\mathbb{C}^2$, with $b$ and $f$ projecting to the `bosonic' and `fermionic' parts of $\mathbb{C}^2$. It is also important to note that the $\mathbb{C}^2$ representation of SUSY \eqref{eq:SUSY-C2} span the space of operators acting on $\mathbb{C}^2$, ${\mathcal Mat}_2(\mathbb{C})$. In fact they are precisely the canonical $e_{ij}=\delta_{ij}$ basis of ${\mathcal Mat}_2(\mathbb{C})$. Thus these elements can be used to generate all operators acting on a $N$-fold tensor product of $\mathbb{C}^2$. Thus substituting these matrices into the above SUSY realizations of the $A$ and $B$ operators we find several constant non-invertible $4\times 4$ Yang-Baxter solutions. However all of these are equivalent to one of the 12 classes of constant non-invertible solutions found by Hietarinta \cite{HIETARINTA-PLA} under the symmetries \eqref{eq:cYBE-gauge}, and \eqref{eq:YBE-discrete-I}-\eqref{eq:YBE-discrete-III}.  Next we will systematically apply this property to realize all the $4\times 4$ constant solutions of Hietarinta using these SUSY algebras. 

\section{The $4\times 4$ local non-hermitian Hamiltonians}
\label{sec:4x4-nhHamiltonians}
Hietarinta's work on the classification of constant $4\times 4$ Yang-Baxter solutions includes 12 classes of non-invertible $Y$ operators. We will denote these classes as $Rp,q$ following Hietarinta \cite{HIETARINTA-PLA}. The integer $p$ denotes the number of independent parameters in the solution and $q$ stands for the number of that solution for a given $p$. So $R2,4$ means the fourth solution with 2 independent parameters. The indexing does not specify the rank of the solution. As we consider only non-invertible $4\times 4$ solutions, the ranks are three, two and one. 

We will study each of the 12 classes separately. In each case we do the following:
\begin{enumerate}
    \item Each class is represented by a $4\times 4$ matrix with a certain number of parameters \cite{HIETARINTA-PLA}. We will write this $4\times 4$ operator in terms of the SUSY generators in \eqref{eq:SUSY-C2}. As the SUSY generators span the space of operators acting on $\mathbb{C}^2$ \eqref{eq:SUSY-C2}, a parameter appears for each non-zero entry of the $4\times 4$ operator. The resulting expression is algebraic as they can be readily extended to an arbitrary dimension $n$ with an appropriate $\mathbb{Z}_2$-grading.
    \item Next we check the conditions under which this algebraic expression satisfies the constant braided or non-braided YBE. We use the algebra of the SUSY generators \eqref{eq:SUSY-relations} to simplify both the sides of the YBE. This imposes some constraints on the parameters of the SUSY realizations of the $4\times 4$ operators. We then verify if these solutions are new or if they fall into one of the 12 classes of non-invertible solutions in \cite{HIETARINTA-PLA} with a continuous gauge transformation 
    \begin{eqnarray}\label{eq:Q}
        Q= \begin{pmatrix}
            g_1 & g_2 \\ g_3 & g_4
        \end{pmatrix}.
    \end{eqnarray}\textbf{We find no new constant non-invertible $4\times 4$ solution by this approach. However since our methods are algebraic we obtain the higher dimensional versions of all these 12 classes. This means we can obtain the $n^2\times n^2$ analogs of these $4\times 4$ solutions. }
    \item Following this we Baxterize each of these solutions by demanding that they satisfy one of the three Baxterization conditions presented in \eqref{eq:3-conditions}. This results in the regular spectral parameter dependent version of these 12 classes.
    \item The local non-hermitian Hamiltonian is then computed for each class in terms of the SUSY generators. The expressions are algebraic and thus can be written down in any dimension. In the qubit representations they are compared with those found in \cite{de2019classifying,Corcoran:2023zax}. This is done by checking if there is a local similarity transformation $Q\otimes Q$ \eqref{eq:Q}, between the Hamiltonian densities. We also take into account the discrete symmetries of the YBE \eqref{eq:YBE-discrete-I}-\eqref{eq:YBE-discrete-III} while checking for equivalences. \textbf{We find two new classes of local Hamiltonians corresponding to the $R0,4$ and $R0,6$ classes.}
\end{enumerate}

\paragraph{Notation :} A note about the notation in the following subsections. In most constant solutions of Hietarinta, the parameter $q$ appears. This should not be confused with the supercharges generating SUSY algebras. The supercharges come with indices and this should clear the confusion.

\subsection{$R2,4$} 
\label{subsec:R24}
\begin{enumerate}
    \item A representative of this rank 2 class \cite{HIETARINTA-PLA}:
    \begin{equation}
R2,4 = \begin{pmatrix}\label{eq:H2,4-hietarinta}
 0 & \left(p^2-k^2\right) (q-k) & \left(p^2-k^2\right) (k+q) & 0 \\
 0 & 0 & 0 & (k+p) \left(q^2-k^2\right) \\
 0 & 0 & 0 & (p-k) \left(q^2-k^2\right) \\
 0 & 0 & 0 & 0 \\
\end{pmatrix}.
\end{equation}
The SUSY realization is
\begin{equation}\label{eq:Y24-SUSY}
    Y^{2,4}_{ij} = \alpha~ b_i q_j+ \beta~ q_i b_j+ \gamma~ q_i f_j+ \delta~ f_i q_j~~~\alpha,\beta,\gamma,\delta\in\mathbb{C}.
\end{equation}
\item Satisfies both the cnbYBE and cbYBE when
\begin{eqnarray}
    & \alpha \gamma(\beta + \delta) = \beta \delta (\alpha+ \gamma),~~~\textrm{cnbYBE} &  \\
    & \alpha \delta(\beta + \gamma) = \beta \gamma (\alpha+ \delta),~~~~\textrm{cbYBE}. &
\end{eqnarray}
The constant $4\times 4$ YBE solutions become
\begin{equation}
      \tilde{Y}^{2,4}_1 = \begin{pmatrix}
 0 & \alpha  & \beta  & 0 \\
 0 & 0 & 0 & \frac{\alpha  \beta  \gamma }{\beta  (\alpha +\gamma )-\alpha  \gamma } \\
 0 & 0 & 0 & \gamma  \\
 0 & 0 & 0 & 0 \\ 
        \end{pmatrix}, \qquad Y^{2,4}_2 = \begin{pmatrix}
 0 & \alpha  & \beta  & 0 \\
 0 & 0 & 0 & \gamma  \\
 0 & 0 & 0 & \frac{\alpha  \beta  \gamma }{\alpha  (\beta +\gamma )-\beta  \gamma } \\
 0 & 0 & 0 & 0 \\
        \end{pmatrix}.
    \end{equation}
The matrix  $\tilde{Y}^{2,4}_1$ falls into the class $R2,4$ under the following gauge transformation
\begin{eqnarray*}
    & \left\{g_2\to 0,g_3\to 0,g_4\to -\frac{8 \alpha  \beta  g_1 \kappa  k^3 (\alpha  (\beta -\gamma )+\beta  \gamma )}{\gamma ^2 (\alpha -\beta )^3}\right\}      & \\
    & \left\{ p\to k-\frac{2 \alpha  \beta  k}{\alpha  \gamma -\beta  \gamma },q\to \frac{2 \alpha  k}{\beta -\alpha }+k\right\}.       &
\end{eqnarray*}
and $Y^{2,4}_2$ falls into the class $R2,4$ via the gauge transformation
\begin{eqnarray*}
    & \left\{g_2\to 0, g_3\to 0, g_4\to -\frac{8 \alpha  \beta  k^3 \kappa g_1 (\alpha  (\beta +\gamma )-\beta  \gamma )}{\gamma ^2 (\alpha -\beta )^3} \right\}   &  \\ 
    &  \left\{p\to k \left(-\frac{2 \alpha  \beta }{\alpha  \gamma -\beta  \gamma }-1\right),q\to \frac{2 \alpha  k}{\beta -\alpha }+k\right\}.    &
\end{eqnarray*}
\item Baxterization condition I yields:
\begin{equation}
    \alpha \gamma + \beta \delta =0,
\end{equation}
This implies we cannot Baxterize the entire $R2,4$ class in this method.
The algebraic expression for the Hamiltonian is given by
\begin{equation}\label{eq:h24-I-SUSY}
    H_{(2,4);I} = c \sum\limits_j \left[ \alpha~ b_j q_{j+1}+ \beta~ q_j b_{j+1}-\frac{\beta  (\alpha +\beta )}{\alpha -\beta }~ q_j f_{j+1}+ \frac{\alpha  (\alpha +\beta )}{\alpha -\beta }~ f_j q_{j+1}\right].
\end{equation}
\item Baxterization condition III fulfilled when:
\begin{equation}
    \alpha \delta + \beta \gamma =0.
\end{equation}
Once again the full parameter space is not Baxterizable in this method.
The algebraic expression for the Hamiltonian is
\begin{equation}
    H_{(2,4);III} = c \sum\limits_j \left[P_{j,j+1} (\alpha~ b_j q_{j+1}+ \beta~ q_j b_{j+1}+ \gamma~ q_j f_{j+1}+ \delta~ f_j q_{j+1})\right],
\end{equation}
subject to the following two constraints
\begin{align*}
    & \gamma = \frac{\alpha  (\alpha +\beta )}{\alpha -\beta },~~\delta = -\frac{\beta  (\alpha +\beta )}{\alpha -\beta }.    
\end{align*}
\item In the qubit representation, the local terms take the form
\begin{eqnarray}
\mathfrak{h}_{(2,4);I}&=&\mathfrak{h}_{(2,4);III}= \begin{pmatrix}
 0 & \alpha  & \beta  & 0 \\
 0 & 0 & 0 & -\frac{\beta  (\alpha +\beta )}{\alpha -\beta } \\
 0 & 0 & 0 & \frac{\alpha  (\alpha +\beta )}{\alpha -\beta } \\
 0 & 0 & 0 & 0 \\
\end{pmatrix}. \nonumber\\
&=& \frac{\alpha \beta}{(\alpha-\beta)}(\sigma^+_j Z_{j+1}-Z_j \sigma^+_{j+1})+\frac{\beta^2}{(\beta-\alpha)} \sigma^+_j+\frac{\alpha^2}{(\alpha-\beta)} \sigma^+_{j+1}.
\end{eqnarray}
Here $\sigma^{\pm} =\frac{1}{2} (X \pm\mathrm{i}Y)$, with $X$, $Y$ and $Z$ being the three Pauli matrices.
The local term $\mathfrak{h}_{(2,4);I(III)}$ is equivalent to non-diagonalizable class-3 model of \cite{de2019classifying}:
\begin{equation}\label{eq:marius-class3}
\begin{pmatrix}
 0 & a_2 & a_3-a_2 & a_5 \\
 0 & a_1 & 0 & a_4 \\
 0 & 0 & -a_1 & a_3-a_4 \\
 0 & 0 & 0 & 0 \\
\end{pmatrix},
\end{equation}
under the gauge transformations
$$\left\{g_3\to 0,a_1\to 0,a_2\to \frac{\alpha  g_4}{g_1 \kappa },a_3\to \frac{g_4 (\alpha +\beta )}{g_1 \kappa },a_4\to \frac{\beta  g_4 (\alpha +\beta )}{g_1 \kappa  (\beta -\alpha )},a_5\to 0\right\}.$$
\end{enumerate}

\subsection{$R1,5$} 
\label{subsec:R15}
\begin{enumerate}
    \item The rank 3 representative is \cite{HIETARINTA-PLA}
    \begin{equation}
R1,5 = \begin{pmatrix}\label{eq:H1,5-hietarinta}
 p+q & 0 & 0 & 0 \\
 0 & q & 0 & q \\
 0 & 0 & p+q & 0 \\
 0 & p & 0 & p \\
\end{pmatrix}.
\end{equation}
Its SUSY realization:
\begin{equation}\label{eq:Y15-SUSY}
    Y^{1,5}_{ij} = \alpha~ q^\dag_i f_j +\mu~ f_i f_j+ \beta~ q^\dag_i q_j+ \nu~f_i q_j+ \gamma~ b_i b_j+\delta~ q_i q^\dag_j.
\end{equation}
The parameters $\alpha$, $\mu$, $\beta$, $\nu$, $\gamma$ and $\delta$ are complex scalars.
\item  Satisfies the cbYBE when
\begin{equation}\label{eq:Y15-c1}
    \gamma = \delta = \mu+ \beta, ~\And~ \nu =\frac{\mu \beta}{\alpha}
\end{equation}
The constant YBE solution in the local $\mathbb{C}^2$ representation becomes
\begin{equation}
 Y^{1,5}= \begin{pmatrix}
 \beta +\mu  & 0 & 0 & 0 \\
 0 & 0 & \beta +\mu  & 0 \\
 0 & \beta  & 0 & \frac{\beta  \mu }{\alpha } \\
 0 & \alpha  & 0 & \mu  \\
    \end{pmatrix},
\end{equation}
which is equivalent to $R1,5$ class itself with the following gauge conditions
$$\left\{g_2\to 0,g_3\to 0,g_4\to \frac{\alpha  g_1}{\mu },p\to \frac{\mu }{\kappa },q\to \frac{\beta }{\kappa }\right\}.$$

\item This satisfies Baxterization condition II, $Y^2 =\eta~ Y$ when $\gamma=\delta = 0$. With this additional constraint \eqref{eq:Y15-SUSY} becomes a rank 1 solution.
With these constraints this solution falls into the subclass of $R1,5$ itself and the special class of rank-2 $(R1,8)^T$ class
$$\begin{pmatrix}
 0 & 0 & 0 & 0 \\
 p & k & 0 & 0 \\
 0 & 0 & 0 & 0 \\
 q & 0 & 0 & 0 \\
\end{pmatrix}$$
with the gauge transformation 
$$\left\{g_1\to 0,g_4\to \frac{\alpha  g_2}{\beta },k\to -\frac{\beta }{\kappa },p\to -\frac{\beta ^2 g_3}{\alpha  g_2 \kappa },q\to 0\right\}.$$
\item The algebraic expression for the Hamiltonian is
\begin{equation}\label{eq:H15;II}
    H_{(1,5);II} = -\frac{c}{\beta} \sum\limits_j (\alpha~q^\dag_j -\beta~ f_j ) (f_{j+1}+\frac{\beta}{\alpha}~q_{j+1}).
\end{equation}
\item In the $\mathbb{C}^2$ representation, the matrix form of the local Hamiltonian density is given by  
\begin{eqnarray}
  \mathfrak{h}_{(1,5);II} &=& \begin{pmatrix}
 0 & 0 & 0 & 0 \\
 0 & 0 & 0 & 0 \\
 0 & \beta  & 0 & -\frac{\beta ^2}{\alpha } \\
 0 & \alpha  & 0 & -\beta  \\
    \end{pmatrix}.\nonumber\\
&=& -\frac{\beta}{4}(\mathbb{1}+Z_jZ_{j+1})+ \beta~ \sigma^-_j \sigma^+_{j+1}+\frac{\alpha}{2}~\sigma^-_j(\mathbb{1}-Z_{j+1}) \nonumber\\& +&\frac{\beta^2}{2\alpha}(Z_j-\mathbb{1})\sigma^+_{j+1}+ \frac{\beta}{4}(Z_j+Z_{j+1}).\nonumber\\
\end{eqnarray}
This local term falls into rank-2 model of \cite{de2019classifying}
\begin{equation}\label{eq:marius-d-model5}
\begin{pmatrix}
 0 & C (2 A-B) & C (2 A+B) & 0 \\
 0 & 2 A-B & 0 & 0 \\
 0 & 0 & 2 A+B & 0 \\
 0 & 0 & 0 & 0 \\
\end{pmatrix},
\end{equation}
with the following gauge transformation
$$\left\{g_1\to -\frac{\alpha  g_3}{\beta },g_4\to 0,B\to -2 A,C\to \frac{\beta  g_2}{\alpha  g_3},\kappa \to -\frac{\beta }{4 A}\right\}.$$
\end{enumerate}

\subsection{$R1,6$} 
\label{subsec:R16}
\begin{enumerate}
    \item The rank 3 representative is \cite{HIETARINTA-PLA} is
    \begin{equation}
R1,6 = \begin{pmatrix}\label{eq:H1,6-hietarinta}
 0 & p & p & 0 \\
 0 & 0 & k & q \\
 0 & k & 0 & q \\
 0 & 0 & 0 & 0 \\
\end{pmatrix}.
\end{equation}
The SUSY version is 
\begin{equation}\label{eq:Y16-SUSY}
    Y^{1,6}_{ij} = \alpha~ b_i f_j+\mu~ f_i b_j+ \beta~ b_i q_j+\nu~ q_i b_j+ \gamma~ q_i f_j+\delta~ f_i q_j.
\end{equation}
The parameters $\alpha$, $\mu$, $\beta$, $\nu$, $\gamma$ and $\delta$ are complex scalars.
\item Satisfies the cbYBE if
    \begin{equation}
        \alpha =\mu, ~ \beta =\nu, ~\And~ \gamma= \delta,
    \end{equation}
which yields identical representation with the class $R1,6$ for local Hilbert space $\mathbb{C}^2$.
\item Satisfies Baxterization condition II when $\beta \gamma = 0$.
Results in two rank 2 solution from here as described below:
\begin{equation}
    Y^{1,6}_{ij;r} =
    \begin{cases}
        \alpha~ (b_i f_j+ f_i b_j)+ \beta~ (b_i q_j+ q_i b_j), ~~\text{when} ~ \gamma=0\\
        \alpha~ (b_i f_j+ f_i b_j)+\gamma~ (q_i f_j+ f_i q_j) , ~~\text{when} ~ \beta=0
    \end{cases}  
    \end{equation}
\item The corresponding algebraic expressions for the local Hamiltonians are given by 
\begin{equation}\label{eq:H-1,6-hamiltonians}
    H_{(1,6);II,1(2)} = 
    \begin{cases}
        \frac{c}{\alpha} \displaystyle\sum\limits_j \left[\alpha~ (b_j f_{j+1}+ f_j b_{j+1})+ \beta~ (b_j q_{j+1}+ q_j b_{j+1})\right] \\
        \frac{c}{\alpha} \displaystyle\sum\limits_j \left[\alpha~ (b_j f_{j+1}+ f_j b_{j+1})+\gamma~ (q_j f_{j+1}+ f_j q_{j+1})\right].
    \end{cases}
\end{equation}
\item In the qubit representation the matrix form of the local terms are
\begin{eqnarray}
\mathfrak{h}_{(1,6);II,1} &=&\begin{pmatrix}
 0 & \beta  & \beta  & 0 \\
 0 & \alpha  & 0 & 0 \\
 0 & 0 & \alpha  & 0 \\
 0 & 0 & 0 & 0 \\
\end{pmatrix} \nonumber\\
&=& \frac{\alpha}{2}(\mathbb{1}-Z_jZ_{j+1})+\frac{\beta}{2}(Z_j \sigma^+_{j+1}+\sigma^+_jZ_{j+1})+\frac{\beta}{2}(\sigma^+_j+\sigma^+_{j+1}), \nonumber\\ 
\mathfrak{h}_{(1,6);II,2}&=& \begin{pmatrix}
 0 & 0 & 0 & 0 \\
 0 & \alpha  & 0 & \beta  \\
 0 & 0 & \alpha  & \beta  \\
 0 & 0 & 0 & 0 \\
\end{pmatrix} \nonumber\\
&=& \frac{\alpha}{2}(\mathbb{1}-Z_jZ_{j+1})-\frac{\beta}{2}(Z_j \sigma^+_{j+1}+\sigma^+_jZ_{j+1})+\frac{\beta}{2}(\sigma^+_j+\sigma^+_{j+1}).
\end{eqnarray}
The matrix $\mathfrak{h}_{(1,6);II,1}$ falls into \eqref{eq:marius-d-model5}
with the following gauge transformation
$$\left\{g_2\to 0,g_3\to 0,A\to \frac{\alpha }{2 \kappa },B\to 0,C\to \frac{\beta  g_4}{\alpha  g_1}\right\},$$
and $\mathfrak{h}_{(1,6);II,2}$ falls into the same category for the gauge transformations
$$\left\{g_3\to 0,g_4\to -\frac{\alpha  g_2}{\beta },A\to \frac{\alpha }{2 \kappa },B\to 0,C\to -\frac{g_2}{g_1}\right\}.$$
\end{enumerate}

\subsection{$R1,7$} 
\label{subsec:R17}
\begin{enumerate}
    \item A rank 2 representative is \cite{HIETARINTA-PLA}:
    \begin{equation}
R1,7 = \begin{pmatrix}\label{eq:H1,7-hietarinta}
 0 & k (k-q) & -k (k+q) & p^2 \\
 0 & 0 & 0 & q (q-k) \\
 0 & 0 & 0 & -q (k+q) \\
 0 & 0 & 0 & 0 \\
\end{pmatrix}.
\end{equation}
The SUSY realization is
\begin{equation}\label{eq:Y17-SUSY}
    Y^{1,7}_{ij} = \alpha~ b_i q_j+ \beta~ q_i b_j+ \gamma~ q_i f_j+ \delta~ f_i q_j+ \mu~ q_i q_j.
\end{equation}
The parameters $\alpha$, $\mu$, $\beta$, $\gamma$ and $\delta$ are complex scalars.
\item Satisfies both the cnbYBE and cbYBE for the same constraints that appear in the $R2,4$ class.
\item Baxterized in two ways
\begin{equation}
   \begin{cases}
     \alpha \delta + \beta \gamma =0, ~~ \text{for}~~ PYPY=0~~\textrm{Condition III}\\
     \alpha \gamma + \beta \delta =0, ~~ \text{for}~~ Y^2=0 ~~\textrm{Condition I}.
    \end{cases}
\end{equation}
\item The algebraic expressions for the Hamiltonians are given by
\begin{eqnarray}\label{eq:h17-SUSY}
    &  H_{(1,7);I} = c \sum\limits_j \left[ \alpha~ b_j q_{j+1}+ \beta~ q_j b_{j+1}-\frac{\beta  (\alpha +\beta )}{\alpha -\beta }~ q_j f_{j+1} \right. & \nonumber \\ & \left. + \frac{\alpha  (\alpha +\beta )}{\alpha -\beta }~ f_j q_{j+1} +\mu~ q_j q_{j+1}\right], &\nonumber \\
    & H_{(1,7);III} = c \sum\limits_j \left[P_{j,j+1} \left(\alpha~ b_j q_{j+1}+ \beta~ q_j b_{j+1}+\frac{\alpha  (\alpha +\beta )}{\alpha -\beta }~ q_j f_{j+1} \right.\right. & \nonumber \\ 
    & \left.\left.-\frac{\beta  (\alpha +\beta )}{\alpha -\beta } ~ f_j q_{j+1} +\mu~ q_j q_{j+1}\right)\right]. &
\end{eqnarray}
\item The matrix forms of the local Hamiltonians in the qubit representation are given by 
\begin{eqnarray}\label{eq:h1,7}
\mathfrak{h}_{(1,7);I} &=& \mathfrak{h}_{(1,7);III} =\begin{pmatrix}
 0 & \alpha  & \beta  & \mu  \\
 0 & 0 & 0 & -\frac{\beta  (\alpha +\beta )}{\alpha -\beta } \\
 0 & 0 & 0 & \frac{\alpha  (\alpha +\beta )}{\alpha -\beta } \\
 0 & 0 & 0 & 0 \\
\end{pmatrix} \nonumber\\
&=& \mu~\sigma^+_j \sigma^+_{j+1}+\frac{\alpha \beta}{(\alpha-\beta)}(\sigma^+_j Z_{j+1}-Z_j \sigma^+_{j+1})+\frac{\beta^2}{(\beta-\alpha)} \sigma^+_j+\frac{\alpha^2}{(\alpha-\beta)} \sigma^+_{j+1}.\nonumber\\
\end{eqnarray}
They fall into the class-3 model \eqref{eq:marius-class3}
under the gauge transformations described below
$$\left\{g_3\to 0,a_1\to 0,a_2\to \frac{\alpha  g_4}{g_1 \kappa },a_3\to \frac{g_4 (\alpha +\beta )}{g_1 \kappa },a_4\to \frac{\beta  g_4 (\alpha +\beta )}{g_1 \kappa  (\beta -\alpha )},a_5\to \frac{g_4^2 \mu }{g_1^2 \kappa }\right\}.$$
\end{enumerate}

\subsection{$R1,8$} 
\label{subsec:R18}
\begin{enumerate}
    \item The rank 2 representative is \cite{HIETARINTA-PLA}
    \begin{equation}\label{eq:H1,8-hietarinta}
R1,8 = \begin{pmatrix}
 0 & p & 0 & q \\
 0 & 0 & 0 & 0 \\
 0 & k & 0 & 0 \\
 0 & 0 & 0 & 0 \\
\end{pmatrix}.
\end{equation}
The SUSY realization
\begin{equation}\label{Y18-SUSY}
    Y^{1,8}_{ij} = \alpha~ b_i f_j+ \beta~ b_i q_j+ \gamma~ q_i q_j~~~\alpha, \beta, \gamma\in\mathbb{C},
\end{equation}
is a solution of cbYBE.
\item Baxterizes under condition II when $\gamma=0$ lowering the rank of the constant solution.
\item Algebraic expression for the local Hamiltonian
\begin{equation}
     H_{(1,8);II} = \frac{c}{\alpha} \sum\limits_j \left[\alpha~ b_j f_{j+1}+ \beta~ b_j q_{j+1}\right].
\end{equation}
\item In the qubit representation the local term takes the form
\begin{eqnarray}
    \mathfrak{h}_{(1,8);II}&=&\begin{pmatrix}
 0 & \beta  & 0 & 0 \\
 0 & \alpha  & 0 & 0 \\
 0 & 0 & 0 & 0 \\
 0 & 0 & 0 & 0 \\
    \end{pmatrix} \nonumber\\
&=& \frac{\alpha}{4}(\mathbb{1}-Z_jZ_{j+1})+\frac{\beta}{2}(Z_j +\mathbb{1})\sigma^+_{j+1} +\frac{\alpha}{4}(Z_j-Z_{j+1}).
\end{eqnarray}
This local term of the Hamiltonian falls into \eqref{eq:marius-d-model5}
with following gauge conditions
$$\left\{g_2\to 0,g_3\to 0,A\to \frac{\alpha }{4 \kappa },B\to -\frac{\alpha }{2 \kappa },C\to \frac{\beta  g_4}{\alpha  g_1}\right\}.$$
\end{enumerate}

\subsection{$R1,9$} 
\label{subsec:R19}
\begin{enumerate}
    \item The rank 2 representative is \cite{HIETARINTA-PLA}:
    \begin{equation}\label{eq:H1,9-hietarinta}
R1,9 = \begin{pmatrix}
 0 & p & 0 & 0 \\
 0 & 0 & 0 & 0 \\
 0 & 0 & 0 & q \\
 0 & 0 & 0 & 0 \\
\end{pmatrix}.
\end{equation}
The SUSY version 
\begin{equation}\label{eq:Y19-SUSY}
    Y^{1,9}_{ij} = (\alpha~ b_i+ \beta~ f_i) q_j~~~\alpha,\beta\in\mathbb{C},
\end{equation}
satisfies both the braided and non-braided YBE's.
\item Entire class Baxterizes under the condition I.
\item Algebraic expression for the local Hamiltonian is
\begin{equation}
     H_{(1,9);I} = c \sum\limits_j \left[(\alpha~ b_j+ \beta~ f_j) q_{j+1}\right].
\end{equation}
\item In the qubit representation the local term becomes
\begin{equation}\label{eq:h1,9}
    \mathfrak{h}_{(1,9);I}=\begin{pmatrix}
 0 & \alpha  & 0 & 0 \\
 0 & 0 & 0 & 0 \\
 0 & 0 & 0 & \beta  \\
 0 & 0 & 0 & 0 \\
    \end{pmatrix} = \left[\frac{(\alpha-\beta)}{2}Z_j +\frac{(\alpha+\beta)}{2}\mathbb{1}\right]\sigma^+_{j+1}.
\end{equation}
This representative does not fall into any of models presented in \cite{de2019classifying}. However, the third discrete transformation of this representative falls into the class-1 nilpotency model of \cite{de2019classifying}
\begin{equation}\label{eq:marius-class1}
\begin{pmatrix}
 0 & a_1 & a_2 & 0 \\
 0 & a_5 & 0 & a_3 \\
 0 & 0 & -a_5 & \frac{a_1 a_3}{a_2} \\
 0 & 0 & 0 & 0 \\
\end{pmatrix}
\end{equation}
under the gauge transformations
$$\left\{g_2\to 0,g_3\to 0,a_1\to 0,a_3\to \frac{a_2 \beta }{\alpha },a_5\to 0,\kappa \to \frac{\alpha  g_4}{a_2 g_1}\right\}.$$
\end{enumerate}

\subsection{$R1,10$} 
\label{subsec:R1,10}
\begin{enumerate}
    \item The rank 2 representative is \cite{HIETARINTA-PLA}:
    \begin{equation}\label{eq:H1,10-hietarinta}
R1,10 = \begin{pmatrix}
 0 & p & 0 & 0 \\
 0 & 0 & 0 & q \\
 0 & 0 & 0 & 0 \\
 0 & 0 & 0 & 0 \\
\end{pmatrix}.
\end{equation}
The SUSY realization
\begin{equation}\label{eq:Y1,10-SUSY}
    Y^{1,10}_{ij} = \alpha~ b_i q_j+ \beta~ q_i f_j~~~\alpha,\beta\in\mathbb{C},
\end{equation}
satisfies both the braided and non-braided YBE. 
\item Baxterizes under condition III.
\item The algebraic expression for the local Hamiltonian is
\begin{equation}
     H_{(1,10);III} = c \sum\limits_j P_{j,j+1}\left[\alpha~ b_j q_{j+1}+ \beta~ q_j f_{j+1}\right].
\end{equation}
\item The qubit representation of the local term in this Hamiltonian
\begin{equation}
  \mathfrak{h}_{(1,10);III}=  \begin{pmatrix}
 0 & \alpha  & 0 & 0 \\
 0 & 0 & 0 & 0 \\
 0 & 0 & 0 & \beta  \\
 0 & 0 & 0 & 0 \\
    \end{pmatrix}= \left[\frac{(\alpha-\beta)}{2}Z_j +\frac{(\alpha+\beta)}{2}\mathbb{1}\right]\sigma^+_{j+1}.
\end{equation}
As in the $R1,9$ class, the third discrete transformation of this representative falls into the class-1 model \eqref{eq:marius-class1} under the gauge transformations
$$\left\{g_2\to 0,g_3\to 0,a_1\to 0,a_3\to \frac{a_2 \beta }{\alpha },a_5\to 0,\kappa \to \frac{\alpha  g_4}{a_2 g_1}\right\}.$$
\end{enumerate}

\subsection{$R1,11$} 
\label{subsec:R1,11}
\begin{enumerate}
    \item The rank 1 representative is \cite{HIETARINTA-PLA}
    \begin{equation}\label{eq:H1,11-hietarinta}
R1,11 = \begin{pmatrix}
 0 & p & q & 0 \\
 0 & 0 & 0 & 0 \\
 0 & 0 & 0 & 0 \\
 0 & 0 & 0 & 0 \\
\end{pmatrix}.
\end{equation}
Its SUSY realization 
\begin{equation}\label{eq:Y1,11-SUSY}
    Y^{1,11}_{ij} = \alpha~ b_i q_j+ \beta~ q_i b_j~~~\alpha,\beta\in\mathbb{C},
\end{equation}
satisfies both braided and non-braided YBE. 
\item Baxterization conditions I and III are satisfied by the entire class.
\item Algebraic expressions for the Hamiltonians are
\begin{eqnarray}
    & H_{(1,11);I} = c \displaystyle\sum\limits_j \left[ \alpha~ b_j q_{j+1}+ \beta~ q_j b_{j+1}\right]   & \\
    &  H_{(1,11);III} = c \displaystyle\sum\limits_j \left[P_{j,j+1}( \alpha~ b_j q_{j+1}+ \beta~ q_j b_{j+1}) \right].&
\end{eqnarray}
\item In the qubit representation the local terms of both these Hamiltonians take the form
\begin{equation}\label{eq:h1,11}
  \mathfrak{h}_{(1,11);I(III)} =\begin{pmatrix}
 0 & \alpha  & \beta  & 0 \\
 0 & 0 & 0 & 0 \\
 0 & 0 & 0 & 0 \\
 0 & 0 & 0 & 0 \\
    \end{pmatrix}= \frac{\alpha}{2}(Z_j+\mathbb{1})\sigma^+_{j+1}+\frac{\beta}{2} \sigma^+_j(\mathbb{1}+Z_{j+1}).
\end{equation}
This local term falls into the class-1 model\eqref{eq:marius-class1} under the gauge transformations
$$\left\{g_2\to 0,g_3\to 0,a_1\to \frac{\alpha  a_2}{\beta },a_3\to 0,a_5\to 0,\kappa \to \frac{\beta  g_4}{a_2 g_1}\right\}.$$
\end{enumerate}

\subsection{$R1,12$} 
\label{subsec:R1,12}
\begin{enumerate}
    \item The rank 1 representative is \cite{HIETARINTA-PLA}
    \begin{equation}\label{eq:H1,12-hietarinta}
R1,12 = \begin{pmatrix}
 0 & 0 & 0 & 0 \\
 0 & q & p & 0 \\
 0 & 0 & 0 & 0 \\
 0 & 0 & 0 & 0 \\
\end{pmatrix}.
\end{equation}
The SUSY realization 
\begin{equation}\label{eq:Y1,12-SUSY}
    Y^{1,12}_{ij} = \alpha~ f_i b_j+ \beta~ q^\dag_i q_j~~~\alpha,\beta\in\mathbb{C},
\end{equation}
satisfies both the braided and non-braided YBE.
\item Condition II Baxterizes the entire class.
\item The algebraic expression for the Hamiltonian becomes
\begin{equation}
     H_{(1,12);II} = \frac{c}{\alpha} \sum\limits_j \left[\alpha~ f_j b_{j+1}+ \beta~ q^\dag_j q_{j+1}\right].
\end{equation}
\item The local Hamiltonian density in the qubit representation
\begin{equation}
    \mathfrak{h}_{(1,12);II}=\begin{pmatrix}
 0 & 0 & 0 & 0 \\
 0 & 0 & 0 & 0 \\
 0 & \beta  & \alpha  & 0 \\
 0 & 0 & 0 & 0 \\
    \end{pmatrix}=\frac{\alpha}{4}(\mathbb{1}-Z_j Z_{j+1})+\beta~\sigma^-_j \sigma^+_{j+1}+ \frac{\alpha}{4}(Z_{j+1}-Z_j),
\end{equation}
falls into the special case of $XXZ$ type model
\begin{equation}
    \begin{pmatrix}
 a_1 & 0 & 0 & 0 \\
 0 & b_1 & c_1 & 0 \\
 0 & c_2 & b_2 & 0 \\
 0 & 0 & 0 & a_1 \\
    \end{pmatrix}
\end{equation}
with the following two gauge transformations
\begin{align*}
\scriptsize{\left\{g_1\to 0,g_4\to 0,a_1\to 0,b_1\to \frac{\alpha }{\kappa },c_1\to \frac{\beta }{\kappa },b_2\to 0,c_2\to 0\right\}},\\
\scriptsize{\left\{g_2\to 0,g_3\to 0,a_1\to 0,b_1\to 0,c_1\to 0,b_2\to \frac{\alpha }{\kappa },c_2\to \frac{\beta }{\kappa }\right\}}.
\end{align*}
\end{enumerate}

\subsection{$R0,4$} 
\label{subsec:R04}
\begin{enumerate}
    \item A rank 3 representative is \cite{HIETARINTA-PLA}
    \begin{equation}\label{eq:H0,4-hietarinta}
R0,4 = \begin{pmatrix}
 1 & 0 & 0 & 0 \\
 0 & 0 & 0 & 1 \\
 0 & 1 & 0 & 0 \\
 0 & 0 & 0 & 1 \\
\end{pmatrix}.
\end{equation}
The SUSY realization of this class is given by 
\begin{equation}\label{eq:Y04-SUSY}
     Y^{04}_{ij} = \alpha (b_i b_j+ f_i f_j+ b_i f_j) + \beta f_i q_j~~~\alpha,\beta\in\mathbb{C}.
\end{equation}
This satisfies cbYBE.
\item The entire class Baxterizes under condition II.
\item The algebraic expression for the Hamiltonian is
\begin{equation}
     H_{(0,4);II} = \frac{c}{\alpha} \sum\limits_j \left[\alpha (b_j b_{j+1}+ f_j f_{j+1}+ b_j f_{j+1}) + \beta~ f_j q_{j+1}\right].
\end{equation}
\item The qubit representation of the local term of the Hamiltonian is
\begin{equation}
   \mathfrak{h}_{(0,4);II} =  \begin{pmatrix}
 \alpha  & 0 & 0 & 0 \\
 0 & \alpha  & 0 & 0 \\
 0 & 0 & 0 & \beta  \\
 0 & 0 & 0 & \alpha  \\
    \end{pmatrix}
   = \frac{\alpha}{4}(3\mathbb{1} + Z_j Z_{j+1})- \frac{\beta}{2}(Z_j-1)\sigma^{+}_{j+1}+\frac{\alpha}{4}(Z_j- Z_{j+1}).
\end{equation}
Note that the last term in this Hamiltonian vanishes on a closed chain or periodic boundary conditions. The model is a deformation of the Ising model. It does not fall in any of the classes of Hamiltonians presented in \cite{de2019classifying} and two additional models (page 12 of \cite{Corcoran:2023zax}) derived from $R$-matrices of difference form.
\end{enumerate}

\subsection{$R0,5$} 
\label{subsec:R05}
\begin{enumerate}
    \item Another rank 3 representative is \cite{HIETARINTA-PLA}
    \begin{equation}\label{eq:H0,5-hietarinta}
R0,5 = \begin{pmatrix}
 1 & 0 & 0 & 0 \\
 0 & 0 & 1 & 0 \\
 0 & 1 & 0 & 0 \\
 0 & 0 & 0 & 0 \\
\end{pmatrix}.
\end{equation}
The SUSY expression is
\begin{equation}\label{eq:Y0,5-SUSY}
    Y^{05}_{ij} = b_i b_j+ b_i f_j + f_i b_j.
\end{equation}
\item Baxterizes with condition II.
\item The algebraic expression for the local Hamiltonian is
\begin{equation}
     H_{(0,5);II} = c \sum\limits_j \left[ b_j b_{j+1}+ b_j f_{j+1} + f_j b_{j+1} \right].
\end{equation}
\item The Hamiltonian density in the qubit representation 
\begin{equation}
   \mathfrak{h}_{(0,5);II} =\begin{pmatrix}
 1 & 0 & 0 & 0 \\
 0 & 1 & 0 & 0 \\
 0 & 0 & 1 & 0 \\
 0 & 0 & 0 & 0 \\
    \end{pmatrix}= \frac{1}{4}(3 \mathbb{1}-Z_jZ_{j+1}+Z_j+Z_{j+1}),
\end{equation}
falls into the special case of the 4-vertex model under the gauge transformation
\begin{align*}
 & \scriptsize{\left\{g_1\to 0,g_4\to 0,a_1\to 0,b_1\to \frac{1}{\kappa },a_2\to \frac{1}{\kappa },b_2\to \frac{1}{\kappa }\right\},} \\
& \scriptsize{   \left\{g_2\to 0,g_3\to 0,a_1\to \frac{1}{\kappa },b_1\to \frac{1}{\kappa },a_2\to 0,b_2\to \frac{1}{\kappa }\right\}}.
\end{align*}
\end{enumerate}

\subsection{$R0,6$} 
\label{subsec:R06}
\begin{enumerate}
    \item The rank 2 representative is \cite{HIETARINTA-PLA}
    \begin{equation}\label{eq:H0,6-hietarinta}
R0,6 = \begin{pmatrix}
 1 & 1 & 1 & 0 \\
 0 & 0 & 0 & 0 \\
 0 & 0 & 0 & 0 \\
 0 & 0 & 0 & 1 \\
\end{pmatrix}.
\end{equation}
The SUSY realization 
\begin{equation}\label{eq:Y06-SUSY}
      Y^{06}_{ij} = \alpha (b_i b_j+ f_i f_j)+ \beta (b_iq_j+ q_i b_j)~~~\alpha,\beta\in\mathbb{C},
\end{equation}
satisfies the cbYBE.
\item Baxterizes under condition II.
\item Algebraic expression for the local Hamiltonian is
\begin{equation}
     H_{(0,6);II} = \frac{c}{\alpha} \sum\limits_j \left[\alpha (b_j b_{j+1}+ f_j f_{j+1})+ \beta (b_jq_{j+1}+ q_j b_{j+1})\right].
\end{equation}
In the $\mathbb{C}^2$ representation the Hamiltonian density
\begin{equation}
  \mathfrak{h}_{(0,6);II} = \begin{pmatrix}
 \alpha  & \beta  & \beta  & 0 \\
 0 & 0 & 0 & 0 \\
 0 & 0 & 0 & 0 \\
 0 & 0 & 0 & \alpha  \\
    \end{pmatrix}
 =\frac{\alpha}{2}(\mathbb{1} + Z_j Z_{j+1})+ \frac{\beta}{2}(\sigma^{+}_j Z_{j+1}+ Z_j \sigma^{+}_{j+1} + \sigma^{+}_j+ \sigma^{+}_{j+1}),
\end{equation}
is not equivalent to any of the classes \cite{de2019classifying} and the two additional models (\cite{Corcoran:2023zax}, p-12) derived from the difference form of the $R$-matrices.
\end{enumerate}

\subsection{Summary of inequivalent non-hermitian Hamiltonians} 
\label{subsec:Summary}
The inequivalent local non-hermitian Hamiltonians are summed up in Table \ref{tab:summary}. We find 9 inequivalent classes. Class $R1,5$ is not included as their $4\times 4$ representation is equivalent by a gauge transformation to the $4\times 4$ representation of $R1,8^T$. The two Hamiltonians obtained in class $R1,6$ \eqref{eq:H-1,6-hamiltonians} are equivalent to each other and so we include just one of them here. 
The class $R1,10$ gives the same Hamiltonian density as $R1,9$ in the $4\times 4$ representation, and so we have not included it. We have also omitted $R2,4$ as it can be obtained from $R1,7$ when $\mu=0$.
 
Note that all the Hamiltonians, except the one obtained from the $R0,5$ class, are `naturally' non-hermitian without the need of making the parameters complex. The local Hamiltonian corresponding to the $R0,5$ class contains only diagonal elements and so this becomes non-hermitian only when the overall constant $c$ is set to a complex number.

\begin{longtable}{ |c|c|c| }
 \hline 
 Class & Algebraic Hamiltonian & $4 \times 4$ Hamiltonian density\\
 \hline \hline
 $R1,6$ & $\frac{c}{\alpha} \displaystyle{\sum_j \left[\alpha~ (b_j f_{j+1}+ f_j b_{j+1})+ \beta~ (b_j q_{j+1}+ q_j b_{j+1})\right]}$ & $\begin{pmatrix}
 0 & \beta  & \beta  & 0 \\
 0 & \alpha  & 0 & 0 \\
 0 & 0 & \alpha  & 0 \\
 0 & 0 & 0 & 0 \\
\end{pmatrix}$\\
 \hline
 $R1,7$ & $\scriptsize{c \sum\limits_j \left[ \alpha~ b_j q_{j+1}+ \beta~ q_j b_{j+1}-\frac{\beta  (\alpha +\beta )}{\alpha -\beta }~ q_j f_{j+1}+ \frac{\alpha  (\alpha +\beta )}{\alpha -\beta }~ f_j q_{j+1} +\mu~ q_j q_{j+1}\right]}$ &  $\begin{pmatrix}
 0 & \alpha  & \beta  & \mu  \\
 0 & 0 & 0 & -\frac{\beta  (\alpha +\beta )}{\alpha -\beta } \\
 0 & 0 & 0 & \frac{\alpha  (\alpha +\beta )}{\alpha -\beta } \\
 0 & 0 & 0 & 0 \\
\end{pmatrix}$ \\
 \hline
 $R1,8$ & $\displaystyle{\frac{c}{\alpha} \sum\limits_j \left[\alpha~ b_j f_{j+1}+ \beta~ b_j q_{j+1}\right]}$ &  $\begin{pmatrix}
 0 & \beta  & 0 & 0 \\
 0 & \alpha  & 0 & 0 \\
 0 & 0 & 0 & 0 \\
 0 & 0 & 0 & 0 \\
    \end{pmatrix}$ \\
 \hline
 $R1,9$ & $\displaystyle{c \sum\limits_j \left[(\alpha~ b_j+ \beta~ f_j) q_{j+1}\right]}$ &  $\begin{pmatrix}
 0 & \alpha  & 0 & 0 \\
 0 & 0 & 0 & 0 \\
 0 & 0 & 0 & \beta  \\
 0 & 0 & 0 & 0 \\
    \end{pmatrix}$ \\
 \hline
 $R1,11$ & $c \displaystyle{\sum\limits_j \left[ \alpha~ b_j q_{j+1}+ \beta~ q_j b_{j+1}\right]}$ &  $\begin{pmatrix}
 0 & \alpha  & \beta  & 0 \\
 0 & 0 & 0 & 0 \\
 0 & 0 & 0 & 0 \\
 0 & 0 & 0 & 0 \\
    \end{pmatrix}$ \\
 \hline
 $R1,12$ & $\displaystyle{\frac{c}{\alpha} \sum\limits_j \left[\alpha~ f_j b_{j+1}+ \beta~ q^\dag_j q_{j+1}\right]}$ & $\begin{pmatrix}
 0 & 0 & 0 & 0 \\
 0 & 0 & 0 & 0 \\
 0 & \beta  & \alpha  & 0 \\
 0 & 0 & 0 & 0 \\
    \end{pmatrix}$  \\
 \hline
 $R0,4$  & $\displaystyle{\frac{c}{\alpha} \sum\limits_j \left[\alpha (b_j b_{j+1}+ f_j f_{j+1}+ b_j f_{j+1}) + \beta f_j q_{j+1}\right]}$ &  $\begin{pmatrix}
 \alpha  & 0 & 0 & 0 \\
 0 & \alpha  & 0 & 0 \\
 0 & 0 & 0 & \beta  \\
 0 & 0 & 0 & \alpha  \\
    \end{pmatrix}$ \\
 \hline
 $R0,5$  & $c \displaystyle\sum\limits_j \left[ b_j b_{j+1}+ b_j f_{j+1} + f_j b_{j+1} \right]$ & $\begin{pmatrix}
 1 & 0 & 0 & 0 \\
 0 & 1 & 0 & 0 \\
 0 & 0 & 1 & 0 \\
 0 & 0 & 0 & 0 \\
    \end{pmatrix}$  \\
 \hline
 $R0,6$  & $\displaystyle{\frac{c}{\alpha} \sum\limits_j \left[\alpha (b_j b_{j+1}+ f_j f_{j+1})+ \beta (b_jq_{j+1}+ q_j b_{j+1})\right]}$ & $\begin{pmatrix}
 \alpha  & \beta  & \beta  & 0 \\
 0 & 0 & 0 & 0 \\
 0 & 0 & 0 & 0 \\
 0 & 0 & 0 & \alpha  \\
    \end{pmatrix}$  \\
 \hline
 \caption{The different inequivalent non-hermitian Hamiltonians obtained in Sec. \ref{sec:4x4-nhHamiltonians}. The algebraic expressions in the second column are written in terms of the elements of the SUSY algebra. Substituting the two dimensional representation \eqref{eq:SUSY-C2} yields the $4\times 4$ form in the third column. To obtain higher dimensional spin chains use the prescription outlined in Sec. \ref{subsec:higherdimensions} on the algebraic expressions of the second column. The parameters $\alpha$, $\mu$, $\beta$, $\nu$, $\gamma$ and $\delta$ are complex scalars.} \label{tab:summary}
\end{longtable}

\section{Features of the algebraic method}
\label{sec:features}
Several results are an immediate consequence of the algebraic nature of the expressions presented, be it the $R$-matrices or the Hamiltonians.
First we see that the regular $R$-matrices obtained in this work satisfy the braided unitarity condition,
\begin{equation}
    \tilde{R}_{ij}(u)\tilde{R}_{ji}(-u) \sim \mathbb{1}.
\end{equation}
It follows immediately from \eqref{eq:R-braided-2},\eqref{eq:R-nonbraided-1} that $\tilde{R}_{ji}(-u) = \tilde{R}^{-1}_{ij}$. 
Next, the expressions for the local Hamiltonians in terms of the SUSY generators enable us to obtain models for any dimension of the local Hilbert space. This is done by choosing an appropriate representation for the SUSY generators in higher dimensions. The algebraic expressions also help us determine when the non-hermitian Hamliltonians are non-diagonalizable and when they are not. It also provides a way to compute the spectrum of some of the low lying states in some cases as we shall demonstrate. 

\subsection{Higher dimensional spin chains}
\label{subsec:higherdimensions}
To construct spin chains in local Hilbert spaces with dimension greater than 2, we first need to find the representations of the SUSY algebra in such spaces. As noted earlier this means that we require a $\mathbb{Z}_2$-grading of $\mathbb{C}^n$ for $n>2$. In dimension 3, there is just one way to do this, namely by splitting the 3 dimensions into a sector of dimension 2 and another sector of dimension 1. It is a matter of choice to call these sectors `bosonic' and `fermionic'. There is additional freedom in choosing the two states that form the `bosonic' sector. These should lead to equivalent systems as the resulting supercharges are related by a rotation. The SUSY generators in one of these cases are written as\footnote{Similar realizations are also interpreted in terms of symmetric inverse semigroups \cite{Padmanabhan_2017}.} 
\begin{eqnarray}\label{eq:SUSYalg-3dim}
    & q = \frac{1}{\sqrt{2}}\begin{pmatrix}
        0 & 1 & 1 \\ 0 & 0 & 0 \\ 0 & 0 & 0
    \end{pmatrix}~;~q^\dag = \frac{1}{\sqrt{2}}\begin{pmatrix}
        0 & 0 & 0 \\ 1 & 0 & 0 \\ 1 & 0 & 0
    \end{pmatrix}      & \nonumber \\
    & b=qq^\dag = \begin{pmatrix}
        1 & 0 & 0 \\ 0 & 0 & 0 \\ 0 & 0 & 0
    \end{pmatrix} ~;~f=q^\dag q=\frac{1}{2}\begin{pmatrix}
        0 & 0 & 0 \\ 0 & 1 & 1 \\ 0 & 1 & 1
    \end{pmatrix}.  &
\end{eqnarray}
Substituting these supercharges into the algebraic expressions for the Hamiltonians in Sec. \ref{sec:4x4-nhHamiltonians} we obtain the corresponding spin 1 models. As an illustration, the Hamiltonian densities of spin 1 integrable models corresponding to the $R0,4$ and $R0,6$ classes are given by
\begin{equation}
 \mathfrak{h}_{(0,4)}^{\text{spin-1}} = \begin{pmatrix}
 \alpha  & 0 & 0 & 0 & 0 & 0 & 0 & 0 & 0 \\
 0 & \frac{\alpha }{2} & \frac{\alpha }{2} & 0 & 0 & 0 & 0 & 0 & 0 \\
 0 & \frac{\alpha }{2} & \frac{\alpha }{2} & 0 & 0 & 0 & 0 & 0 & 0 \\
 0 & 0 & 0 & 0 & \frac{\beta }{2 \sqrt{2}} & \frac{\beta }{2 \sqrt{2}} & 0 & \frac{\beta }{2 \sqrt{2}} & \frac{\beta }{2 \sqrt{2}} \\
 0 & 0 & 0 & 0 & \frac{\alpha }{4} & \frac{\alpha }{4} & 0 & \frac{\alpha }{4} & \frac{\alpha }{4} \\
 0 & 0 & 0 & 0 & \frac{\alpha }{4} & \frac{\alpha }{4} & 0 & \frac{\alpha }{4} & \frac{\alpha }{4} \\
 0 & 0 & 0 & 0 & \frac{\beta }{2 \sqrt{2}} & \frac{\beta }{2 \sqrt{2}} & 0 & \frac{\beta }{2 \sqrt{2}} & \frac{\beta }{2 \sqrt{2}} \\
 0 & 0 & 0 & 0 & \frac{\alpha }{4} & \frac{\alpha }{4} & 0 & \frac{\alpha }{4} & \frac{\alpha }{4} \\
 0 & 0 & 0 & 0 & \frac{\alpha }{4} & \frac{\alpha }{4} & 0 & \frac{\alpha }{4} & \frac{\alpha }{4} \\
    \end{pmatrix}, \quad
 \mathfrak{h}_{(0,6)}^{\text{spin-1}} = \begin{pmatrix}
 \alpha  & \frac{\beta }{\sqrt{2}} & \frac{\beta }{\sqrt{2}} & \frac{\beta }{\sqrt{2}} & 0 & 0 & \frac{\beta }{\sqrt{2}} & 0 & 0 \\
 0 & 0 & 0 & 0 & 0 & 0 & 0 & 0 & 0 \\
 0 & 0 & 0 & 0 & 0 & 0 & 0 & 0 & 0 \\
 0 & 0 & 0 & 0 & 0 & 0 & 0 & 0 & 0 \\
 0 & 0 & 0 & 0 & \frac{\alpha }{4} & \frac{\alpha }{4} & 0 & \frac{\alpha }{4} & \frac{\alpha }{4} \\
 0 & 0 & 0 & 0 & \frac{\alpha }{4} & \frac{\alpha }{4} & 0 & \frac{\alpha }{4} & \frac{\alpha }{4} \\
 0 & 0 & 0 & 0 & 0 & 0 & 0 & 0 & 0 \\
 0 & 0 & 0 & 0 & \frac{\alpha }{4} & \frac{\alpha }{4} & 0 & \frac{\alpha }{4} & \frac{\alpha }{4} \\
 0 & 0 & 0 & 0 & \frac{\alpha }{4} & \frac{\alpha }{4} & 0 & \frac{\alpha }{4} & \frac{\alpha }{4} \\
    \end{pmatrix}.
\end{equation}
For 4 and higher dimensions we can write down more inequivalent supercharges corresponding to the chosen $\mathbb{Z}_2$-grading. For example in dimension 4 we have two choices for the $\mathbb{Z}_2$-grading. We can either spilt the space into two sectors of dimension 2 each or we can split the space into sectors with one having dimension 3 and the other dimension 1. This corresponds to the number of non-zero bi-partitions of the
integer 4. The supercharges corresponding to these two gradings are given by
\begin{eqnarray}\label{eq:SUSYgen-4dim}
    q_{(2,2)} = \frac{1}{\sqrt{2}}\begin{pmatrix}
        0 & 0 & 1 & 1 \\ 0 & 0 & 1 & 1 \\ 0 & 0 & 0 & 0 \\ 0 & 0 & 0 & 0
    \end{pmatrix}~~;~~q_{(3,1)} = \frac{1}{\sqrt{3}}\begin{pmatrix}
        0 & 1 & 1 & 1 \\ 0 & 0 & 0 & 0 \\ 0 & 0 & 0 & 0 \\ 0 & 0 & 0 & 0
    \end{pmatrix}.
\end{eqnarray}
The remaining elements of the SUSY algebra are generated using these expressions. Note that the $2\times 2$ block in $q_{(2,2)}$, $\begin{pmatrix}
    1 & 1 \\ 1& 1
\end{pmatrix}$ can be replaced by any other $2\times 2$ matrix as it does not alter the fact that the resulting matrix is still nilpotent. If the different $2\times 2$ matrices are inequivalent then the resulting supercharges are inequivalent as well leading to inequivalent models for a given grading. We will study them more systematically elsewhere. 

These supercharges lead to spin $\frac{3}{2}$ models when they are substituted into the local Hamiltonians obtained in Sec. \ref{sec:4x4-nhHamiltonians}. For example we have
\begin{equation}
\mathfrak{h}_{(1,7)}^{\text{spin-3/2}}=\frac{1}{\sqrt{2}}  \begin{pmatrix}
 0 & 0 & \alpha  & \alpha  & 0 & 0 & \alpha  & \alpha  & \beta  & \beta  & \frac{\mu }{\sqrt{2}} & \frac{\mu }{\sqrt{2}} & \beta  & \beta  & \frac{\mu }{\sqrt{2}} & \frac{\mu }{\sqrt{2}} \\
 0 & 0 & \alpha  & \alpha  & 0 & 0 & \alpha  & \alpha  & \beta  & \beta  & \frac{\mu }{\sqrt{2}} & \frac{\mu }{\sqrt{2}} & \beta  & \beta  & \frac{\mu }{\sqrt{2}} & \frac{\mu }{\sqrt{2}} \\
 0 & 0 & 0 & 0 & 0 & 0 & 0 & 0 & 0 & 0 & -\frac{\beta  (\alpha +\beta )}{\alpha -\beta } & -\frac{\beta  (\alpha +\beta )}{\alpha -\beta } & 0 & 0 & -\frac{\beta  (\alpha +\beta )}{\alpha -\beta } & -\frac{\beta  (\alpha +\beta )}{\alpha -\beta } \\
 0 & 0 & 0 & 0 & 0 & 0 & 0 & 0 & 0 & 0 & -\frac{\beta  (\alpha +\beta )}{\alpha -\beta } & -\frac{\beta  (\alpha +\beta )}{\alpha -\beta } & 0 & 0 & -\frac{\beta  (\alpha +\beta )}{\alpha -\beta } & -\frac{\beta  (\alpha +\beta )}{\alpha -\beta } \\
 0 & 0 & \alpha  & \alpha  & 0 & 0 & \alpha  & \alpha  & \beta  & \beta  & \frac{\mu }{\sqrt{2}} & \frac{\mu }{\sqrt{2}} & \beta  & \beta  & \frac{\mu }{\sqrt{2}} & \frac{\mu }{\sqrt{2}} \\
 0 & 0 & \alpha  & \alpha  & 0 & 0 & \alpha  & \alpha  & \beta  & \beta  & \frac{\mu }{\sqrt{2}} & \frac{\mu }{\sqrt{2}} & \beta  & \beta  & \frac{\mu }{\sqrt{2}} & \frac{\mu }{\sqrt{2}} \\
 0 & 0 & 0 & 0 & 0 & 0 & 0 & 0 & 0 & 0 & -\frac{\beta  (\alpha +\beta )}{\alpha -\beta } & -\frac{\beta  (\alpha +\beta )}{\alpha -\beta } & 0 & 0 & -\frac{\beta  (\alpha +\beta )}{\alpha -\beta } & -\frac{\beta  (\alpha +\beta )}{\alpha -\beta } \\
 0 & 0 & 0 & 0 & 0 & 0 & 0 & 0 & 0 & 0 & -\frac{\beta  (\alpha +\beta )}{\alpha -\beta } & -\frac{\beta  (\alpha +\beta )}{\alpha -\beta } & 0 & 0 & -\frac{\beta  (\alpha +\beta )}{\alpha -\beta } & -\frac{\beta  (\alpha +\beta )}{\alpha -\beta } \\
 0 & 0 & 0 & 0 & 0 & 0 & 0 & 0 & 0 & 0 & \frac{\alpha  (\alpha +\beta )}{\alpha -\beta } & \frac{\alpha  (\alpha +\beta )}{\alpha -\beta } & 0 & 0 & \frac{\alpha  (\alpha +\beta )}{\alpha -\beta } & \frac{\alpha  (\alpha +\beta )}{\alpha -\beta } \\
 0 & 0 & 0 & 0 & 0 & 0 & 0 & 0 & 0 & 0 & \frac{\alpha  (\alpha +\beta )}{\alpha -\beta } & \frac{\alpha  (\alpha +\beta )}{\alpha -\beta } & 0 & 0 & \frac{\alpha  (\alpha +\beta )}{\alpha -\beta } & \frac{\alpha  (\alpha +\beta )}{\alpha -\beta } \\
 0 & 0 & 0 & 0 & 0 & 0 & 0 & 0 & 0 & 0 & 0 & 0 & 0 & 0 & 0 & 0 \\
 0 & 0 & 0 & 0 & 0 & 0 & 0 & 0 & 0 & 0 & 0 & 0 & 0 & 0 & 0 & 0 \\
 0 & 0 & 0 & 0 & 0 & 0 & 0 & 0 & 0 & 0 & \frac{\alpha  (\alpha +\beta )}{\alpha -\beta } & \frac{\alpha  (\alpha +\beta )}{\alpha -\beta } & 0 & 0 & \frac{\alpha  (\alpha +\beta )}{\alpha -\beta } & \frac{\alpha  (\alpha +\beta )}{\alpha -\beta } \\
 0 & 0 & 0 & 0 & 0 & 0 & 0 & 0 & 0 & 0 & \frac{\alpha  (\alpha +\beta )}{\alpha -\beta } & \frac{\alpha  (\alpha +\beta )}{\alpha -\beta } & 0 & 0 & \frac{\alpha  (\alpha +\beta )}{\alpha -\beta } & \frac{\alpha  (\alpha +\beta )}{\alpha -\beta } \\
 0 & 0 & 0 & 0 & 0 & 0 & 0 & 0 & 0 & 0 & 0 & 0 & 0 & 0 & 0 & 0 \\
 0 & 0 & 0 & 0 & 0 & 0 & 0 & 0 & 0 & 0 & 0 & 0 & 0 & 0 & 0 & 0 \\
    \end{pmatrix}
\end{equation}

This procedure can easily be extended to higher dimensions. The number of gradings for $n>2$ scales as $\lfloor \frac{n}{2}\rfloor$. Thus the algebraic method is flexible in constructing spin chains in all dimensions.

\subsection{Spectrum of the non-hermitian Hamiltonians} 
\label{subsec:spectrumstudy}
In general, non-hermitian Hamiltonians are expected to have a complex spectrum. They may be diagonalizable or not. When they are non-diagonalizable their eigenvectors do not form a complete set. In these cases several eigenvectors of a degenerate eigenvalue coincide. These eigenvalues are called {\it exceptional points} in continuous systems and has many interesting properties \cite{Minganti_2019}. In the discrete setup of spin chains such non-diagonalizable Hamiltonians have appeared in the context of $\mathcal{N}=4$ super Yang-Mills theory \cite{Ipsen_2019} where they are called {\it eclectic spin chains}. Though they are integrable, the usual methods of the algebraic Bethe ansatz do not apply, thus making it a testing ground for the development of new theoretical techniques in integrability. Generalizations of these spin chains have also been considered from a non-gauge theoretic point of view in \cite{Ahn_2021}, where they are constructed from deformed permutation operators. It is thus of interest to classify the Hamiltonians of Table \ref{tab:summary} according to whether they are diagonalizable or not. 

The algebraic forms of the Hamiltonians given in \eqref{eq:H-braided-1}, \eqref{eq:H-nonbraided-1} already show that the local Hamiltonian densities are non-diagonalizable as they are derived from the Baxterization conditions I $(Y^2=0)$ and III $(P\tilde{Y}P\tilde{Y}=0)$ \eqref{eq:3-conditions}. This shows that the local terms of the two Hamiltonians are nilpotent implying that their only eigenvalue is 0. Such matrices will be defective and can only be reduced to a Jordan form. On the other hand the local Hamiltonian density in \eqref{eq:H-braided-2}, obtained from Baxterization condition II $(Y^2=\eta Y)$, shows that these Hamitlonians can be diagonalizable. Here too the algebraic expressions {\it via} the SUSY generators will aid us in settling this problem, thus illustrating the advantage of this formalism. 

From the above arguments the Hamiltonian densities corresponding to the classes- $R1,7, R1,9, R1,11$ are non-diagonalizable (See Table \ref{tab:summary}). The remaining Hamiltonians have diagonalizable local terms. These are easily seen using the SUSY algebra realizations. This does not however, conclusively determine if the total Hamiltonians are diagonalizable or not. This requires more study but we will still consider one of the non-diagonalizable Hamiltonian densities, namely $\mathfrak{h}_{(1,11);I}$. Denote the basis states of $\mathbb{C}^2$ as
$$ \ket{u}=\begin{pmatrix}
    1 \\ 0 
\end{pmatrix}~;~\ket{d}=\begin{pmatrix}
    0 \\ 1 
\end{pmatrix}. $$
Then using the qubit representation of the SUSY generator in \eqref{eq:SUSY-C2}, we see that any product state with consecutive $\ket{u}$'s or consecutive $\ket{d}$'s are annihilated by the local Hamiltonian density. On the other hand, the other two combinations of consecutive symbols $\ket{ud}$ and $\ket{du}$, are converted into the state $\ket{uu}$. This property extends to the global spin chain as well. It is not hard to see that this system has only two zero modes, the two ferromagnetic states
$$ \ket{z_u}=\ket{u_1u_2\cdots u_N}~;~\ket{z_d}=\ket{d_1d_2\cdots d_N}.  $$
The system is frustration-free as both the local and global Hamiltonians kill the above states. The other states contain both $\ket{u}$'s and $\ket{d}$'s. The Hamiltonian maps the state containing $k$ ($k<N$) number of $\ket{d}$'s into one with $k-1$ number of $\ket{d}$'s. Thus states containing both $\ket{u}$'s and $\ket{d}$'s can never become eigenstates of this Hamiltonian. 

Furthermore the time evolution of a non-diagonalizable system is quite different from that of a diagonalizable one. The Jordan form of the non-diagonalizable Hamiltonian is written in terms of upper triangular blocks corresponding to the different eigenvalues $\lambda$. Each such block takes the form 
\begin{eqnarray}
    J_\lambda = \lambda~\mathbb{1} + M_\lambda.
\end{eqnarray}

The evolution operator for that block then becomes 
\begin{eqnarray}
    e^{-\mathrm{i}J_\lambda t} = e^{-\mathrm{i}\lambda t}\sum\limits_{k=0}^{p-1}~\frac{(-\mathrm{i}t)^k}{k!}M^k~~;~~M^p=0.
\end{eqnarray}
This shows that for non-diagonalizable Hamiltonians, the time evolution suppresses the exponential evolution with a polynomial function in time.

A preliminary look at the $R1,11$ Hamiltonian density $\mathfrak{h}_{(1,11);I}$\eqref{eq:h1,11} shows that it can be written as three Jordan blocks: two blocks of size $1$ and one block of size $2$,
$$J_0^{(1)}=(0),~~ J_0^{(2)}=(0), ~~ J_0^{(3)}=\begin{pmatrix}
    0 & 1\\
    0 & 0
\end{pmatrix}.$$
The final expression of Jordan canonical form for $\mathfrak{h}_{(1,11);I}$ is given by
$$J^{(1,11)}= J_0^{(1)} \oplus J_0^{(2)} \oplus J_0^{(3)}.$$
As the only eigenvalue of this matrix is 0, the time evolution of this system is in fact linear in time with no exponential part. However this is the result for the Hamiltonian density. We still need to obtain the Jordan form of the full Hamiltonian. We notice that the full Hamiltonian becomes nilpotent at an exponent that scales with the number of sites $N$. For example at $N=3$, $H^4=0$. Thus in this case the time evolution is polynomial in $t$. Clearly as $N$ becomes very large the time evolution begins to approximate that of the usual exponential function.

The boost operator formalism helps us derive higher order charges for this system. The expression of \textit{spin chain boost operator} \cite{loebbert2016lectures} is given by 
\begin{equation}
    \mathcal{B}= \sum_{j=-\infty}^{\infty} j~ \mathfrak{h}_{j,j+1}.
\end{equation}
We can generate the local higher order charges in a recursive manner
\begin{equation}
    \mathfrak{Q}_{r+1} \sim [\mathcal{B}, \mathfrak{Q}_r].
\end{equation}
For example, the third and fourth order interaction terms for the local Hamiltonians obtained from the $R1,11$ class 
\begin{align}
    \mathfrak{Q}_{j,j+1,j+2}& \sim \beta^2~ q_j q_{j+1} b_{j+2}-\alpha^2~ b_j q_{j+1} q_{j+2}\\
    \mathfrak{Q}_{j,j+1,j+2,j+3}& \sim \beta^3~ q_j q_{j+1} q_{j+2} b_{j+3}-\alpha^3~ b_j q_{j+1} q_{j+2} q_{j+3}.
\end{align}
With some effort it can be seen that these operators are indeed conserved as they commute with the Hamiltonian obtained from the $R1,11$ class. This also indicates that the boost operator method goes through for this non-hermitian system. 

All of this makes the $R1,11$ class a source of an interesting non-hermitian ferromagnetic system that is non-diagonalizable. We expect similar interesting behavior for the other non-hermitian systems as well. A more detailed analysis of all these cases is reserved future publications.

\section{Conclusion}
\label{sec:conclusion}
In this work we have constructed three regular $R$-matrices that satisfy the spectral parameter-dependent Yang-Baxter equations in additive form. Two of them, \eqref{eq:R-braided-1} and \eqref{eq:R-braided-2}, satisfy the braided form of the YBE, whereas one of them, \eqref{eq:R-nonbraided-1}, satisfies the non-braided form [standard or more commonly the $RTT$-form] of the YBE. These solutions are algebraic [representation independent], in the sense that they do not depend on the dimension of the local Hilbert space in which we are working. Thus these $R$-matrices provide solutions in all dimensions when the right representation is chosen. 

These three solutions depend on a constant Yang-Baxter solution $Y$, that can be either invertible or non-invertible. They have to satisfy at least one of the three Baxterization conditions presented in \eqref{eq:3-conditions} for the construction of the spectral parameter dependent version to go through. It is important to note that this is not an exhaustive set of regular $R$-matrices as there are possible extensions of the Baxterization techniques used in this work. We know that this should in fact be true as in a few cases, the technique used did not Baxterize an entire Hietarinta class. It is possible to find more algebraic ans\"{a}tze to Baxterize the entire Hietarinta class.
They could possibly produce new classes of regular solutions. We will consider a more systematic treatment along these lines in the future.

Our technique is tested on the $4\times 4$ constant, non-invertible solutions of Hietarinta \cite{HIETARINTA-PLA}. The 12 solutions are first made algebraic by writing them in terms of SUSY generators as explained in Sec. \ref{subsec:SUSYrealizations}. These give rise to 9 inequivalent local Hamiltonians that are summed up in Table \ref{tab:summary}. They are compared with the solutions of \cite{de2019classifying, Corcoran:2023zax}. We find two new classes coming from the Hietarinta classes $R0,4$ and $R0,6$. In addition to all of them being non-hermitian some are non-diagonalizable. This is surveyed in Sec. \ref{subsec:spectrumstudy}. The algebraic expressions for each one of them allows us to construct the higher dimensional spin chains corresponding to all these solutions. This is described in Sec. \ref{subsec:higherdimensions}. 

There are a few other interesting directions to pursue:
\begin{enumerate}
    \item The regular $R$-matrices for non-hermitian integrable systems while convenient to produce conserved quantities may still not help to solve the model {\it via} algebraic Bethe ansatz. The solutions for these models require new techniques which require further exploration.
    \item We have seen that the non-diagonalizable classes have polynomial time evolution. These would be easier to simulate on a quantum computer. Thus finding the integrable quantum circuits \cite{Sopena_2022,Vanicat2017IntegrableTL,aleiner2021bethe} corresponding to these Hamiltonians would be a worthy pursuit.
    \item We can repeat this analysis for constant invertible $Y$ operators. One would expect hermitian integrable systems in this case. It would be interesting to see if these methods produce new classes of $R$-matrices and local Hamiltonians in this case.
    \item Non-hermitian Hamiltonians can model dissipative systems, open systems or even systems out-of-equilibrium. It would be interesting to interpret the spin chains presented here in these contexts as well.
\end{enumerate}

\section*{Acknowledgments}

VK is funded by the U.S. Department of Energy, Office of Science, National Quantum Information Science Research Centers, Co-Design Center for Quantum Advantage ($C^2QA$) under Contract No. DE-SC0012704.

\bibliographystyle{acm}
\normalem
\bibliography{refs}

\end{document}